\documentclass{article}

\usepackage{arxiv}

\usepackage[utf8]{inputenc} 
\usepackage[T1]{fontenc}    
\usepackage{hyperref}       
\usepackage{url}            
\usepackage{booktabs}       
\usepackage{amsfonts}       
\usepackage{nicefrac}       
\usepackage{microtype}      
\usepackage{lipsum}
\usepackage{graphicx}
\usepackage{amsmath} 
\usepackage{ stmaryrd } 
\usepackage{pythonhighlight} 
\graphicspath{ {./images/} }

\usepackage[square,sort,comma,numbers]{natbib} 
\usepackage{pifont} 

\usepackage{caption, subcaption} 
\usepackage{graphicx} 
\usepackage{pgfplots} 
\pgfplotsset{compat=1.17}

\title{Mixed precision in Graphics Processing Unit}
\author{
 Quentin Gallouédec\\
  École Centrale de Lyon\\
  \texttt{quentin.gallouedec@ec-lyon.fr} \\
}

\begin{document}
\maketitle
\begin{abstract}
Modern graphics computing units (GPUs) are designed and optimized to perform highly parallel numerical calculations. This parallelism has enabled (and promises) significant advantages, both in terms of energy performance and calculation.

In this document, we take stock of the different applications of mixed precision. We recall the standards currently used in the overwhelming majority of systems in terms of numerical computation. We show that the mixed precision which decreases the precision at the input of an operation does not necessarily decrease the precision of its output. We show that this previous principle allows its transposition into one of the branches that most needs computing power: machine learning. The use of fixed point numbers and half-precision are two very effective ways to increase the learning ability of complex neural networks. Mixed precision still requires the use of suitable hardware, failing which the calculation time could on the contrary be lengthened. The NVIDIA Tensor Core that is found among others in their Tesla V100 range, is an example of implementation at the hardware level of mixed precision. On the other hand, by abandoning the traditional von Neumann model, mixed precision can also be transposed to a lower level of abstraction, using phase change memories.
\end{abstract}

\section{Introduction}

Low-precision floating point numbers use fewer bits than high-precision floating point numbers. As a result, they are subject to larger rounding errors.
Therefore, the error caused by rounding can have a large influence on the total error. Some algorithms that use simple precision could not handle the error induced by this decrease in precision. Nevertheless, in many cases, it would be very beneficial to reduce the precision of floating numbers to gain both speed and power. Clustering or graph ranking algorithms or the training of dense neural networks are some examples.
The use of mixed precision may well be a solution to reduce size, power consumption, weight and speed in many computer and electronic applications.

In the first part of this paper, we will first study the IEEE 754 standard that defines the format of the binary representation of real numbers. We will then deduce some theoretical principles fundamental to the use of mixed precision. In the second part, we will study one of the applications that can benefit the most from the development of mixed precision: machine learning. We will see how the chosen binary representation is at the heart of the performance of its algorithms. Finally, in a last part, we will see the current research axes to dimension the hardware to take full advantage of the application of mixed precision.

\section{IEEE 754 : Standard for Floating-Point Arithmetic}

The vast majority of computers use \textit{floating-point arithmetic} to represent real numbers.
The technical standard for floating-point arithmetic was established in 1985 by the IEEE Standard for Floating-Point Arithmetic (IEEE 754)\cite{ieeefloatingpoint}. Since memories use the binary form, there is necessarily a loss of precision in the transition from real numbers to floating point numbers.

\subsection{Single-precision floating-point format}

The single-precision floating-point format uses 32 bits of memory. This is why it is called \texttt{float32}. 
A half-precision floating-point number requires 16 bits of memory, and a double-precision floating-point number uses 64 bits of memory.
Throughout this paper, we will use \texttt{float16}, \texttt{float32} and \texttt{float64} to refer to a half-, single- and double-precision floating-point number respectively.

Here is the binary representation of a \texttt{float32}. The encoding principle is similar for the encoding of \texttt{float16} and \texttt{float64}

\[
\underbrace{
\texttt{
X}}_{\text{
sign (1 bits)}}~\underbrace{\texttt{XXXXXXX}}_{\text{exponent (8 bits)}}~\underbrace{\texttt{XXXXXXX} \cdots \texttt{XXXXXXX}}_{\text{fraction (23 bits)}}
\]

\noindent
Next, let's index the binary values like this.
\[
\underset{b_{31}}{\texttt{X}} \underset{b_{30}}{\texttt{X}} \underset{b_{29}}{\texttt{X}} \cdots \underset{b_{1}}{\texttt{X}} \underset{b_{0}}{\texttt{X}}
\]

\noindent
Let us denote by $\mathbb{B}^{32}$ the set of 32 boolean sequences.
Thus, $(b_i)_{\mathbb{N}_{<32}} \in \mathbb{B}^{32}$ is the binary representation of the value.

\noindent
Let us set the following values :
\begin{equation}
\label{eq:s_def}
	s = (-1)^{b_{31}}
\end{equation}

\begin{equation}
\label{eq:e_def}
	e = \displaystyle\sum_{i=0}^{7} b_{23+i} 2^i
\end{equation}

\begin{equation}
\label{eq:f_def}
	f = \displaystyle\sum_{i=1}^{23} b_{i-23} 2^{-i}
\end{equation}

Let us denote by $\bar{\mathbb{R}}$ the set of real numbers to which we add the $\{-\infty, +\infty\}$.
Thus, the function $\mathcal{F}^{32}:\mathbb{B}^{32}\longrightarrow \bar{\mathbb{R}}$ that associates the binary representation with its corresponding real number is\footnote{You might notice that the function is not an injection, since $0$ and $\pm \infty$ can be coded in several different ways.} :

\begin{equation}
\mathcal{F}^{32}: (b_i)_{\mathbb{N}_{<32}} \mapsto\\ 
	\begin{cases}
		s  \times 2^{-127} \times f \quad \text{if} \quad e = 0\\
		s  \times 2^{e-127} \times \left( 1 + f \right) \quad \text{if} \quad e \in
		\llbracket 1, 254\rrbracket \\
		s \times +\infty \quad \text{if} \quad e = 255\\
	\end{cases}
\end{equation}

\noindent
For $e = 0$ the numbers are called \textit{sub-normal numbers}.

\paragraph{Examples}
\begin{align*}
    \texttt{0 00000000 00000000000000000000000} &= (-1)^0 \times 2^{-127} \times 0 = 0\\
    \texttt{1 10000000 00000000000000000000000} &= -1 \times 2^{128-127} \times (1+0) = -2\\
    \texttt{0 10000011 10000000000000000000000} &= 1 \times         2^{131-127} \times (1+2^{-1}) = 12\\
    \texttt{0 01111111 00000000000000000000000} &= 1 \times 2^{127-127} \times (1+0) = 1\\
    \texttt{0 01111111 00000000000000000000001} &= 1+2^{-23} \approx 1.00000011920929
\end{align*}

We denote $F^{32}$ as the image of $\mathcal{F}^{32}$ under $\mathbb{B}^{32}$. The density $F^{32}$ is not uniform over $\mathbb{R}$.
But, in every interval between two power of 2, there are the exact same amount of numbers.

\begin{equation}
    \forall n \in \llbracket -126, 127 \rrbracket \quad  \# F^{32}_{[2^n, 2^{n+1}[}  = 2^{23}
\end{equation}

Where $ F^{32}_{[2^n, 2^{n+1}[} = F^{32} \cap [2^n, 2^{n+1}[$

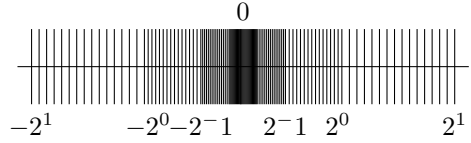
\begin{figure}[ht]
	\centering
	\begin{tikzpicture}
		\draw (-3,0) -- (3,0);
		\draw (0,-0.5) -- (0,0.5) node[anchor=south] {$0$};
		\draw (-0.03125,0.5) -- (-0.03125,-0.5);
		\draw (-0.0125,0.5) -- (-0.0125,-0.5);
		\draw (-0.025,0.5) -- (-0.025,-0.5);
		\draw (-0.037500,0.5) -- (-0.037500,-0.5);
		\draw (-0.05,0.5) -- (-0.05,-0.5);
		\draw (-0.0625,0.5) -- (-0.0625,-0.5);
		\draw (-0.07500,0.5) -- (-0.07500,-0.5);
		\draw (-0.08750,0.5) -- (-0.08750,-0.5);
		\draw (-0.1,0.5) -- (-0.1,-0.5);
		\draw (-0.1125,0.5) -- (-0.1125,-0.5);
		\draw (-0.125,0.5) -- (-0.125,-0.5);
		\draw (-0.1375,0.5) -- (-0.1375,-0.5);
		\draw (-0.15,0.5) -- (-0.15,-0.5);
		\draw (-0.1625,0.5) -- (-0.1625,-0.5);
		\draw (-0.17500,0.5) -- (-0.17500,-0.5);
		\draw (-0.1875,0.5) -- (-0.1875,-0.5);
		\draw (-0.2125,0.5) -- (-0.2125,-0.5);
		\draw (-0.2375,0.5) -- (-0.2375,-0.5);
		\draw (-0.2625,0.5) -- (-0.2625,-0.5);
		\draw (-0.2875,0.5) -- (-0.2875,-0.5);
		\draw (-0.3125,0.5) -- (-0.3125,-0.5);
		\draw (-0.3375,0.5) -- (-0.3375,-0.5);
		\draw (-0.36250,0.5) -- (-0.36250,-0.5);
		\draw (-0.3875,0.5) -- (-0.3875,-0.5);
		\draw (-0.4125,0.5) -- (-0.4125,-0.5);
		\draw (-0.4375,0.5) -- (-0.4375,-0.5);
		\draw (-0.4625,0.5) -- (-0.4625,-0.5);
		\draw (-0.48750,0.5) -- (-0.48750,-0.5);
		\draw (-0.5125,0.5) -- (-0.5125,-0.5);
		\draw (-0.5375,0.5) -- (-0.5375,-0.5);
		\draw (-0.5625,0.5) -- (-0.5625,-0.5) node[anchor=north] {$-2^-1$};
		\draw (-0.6125,0.5) -- (-0.6125,-0.5);
		\draw (-0.6625,0.5) -- (-0.6625,-0.5);
		\draw (-0.7125,0.5) -- (-0.7125,-0.5);
		\draw (-0.7625,0.5) -- (-0.7625,-0.5);
		\draw (-0.8125,0.5) -- (-0.8125,-0.5);
		\draw (-0.8625,0.5) -- (-0.8625,-0.5);
		\draw (-0.9125,0.5) -- (-0.9125,-0.5);
		\draw (-0.9625,0.5) -- (-0.9625,-0.5);
		\draw (-1.0125,0.5) -- (-1.0125,-0.5);
		\draw (-1.0625,0.5) -- (-1.0625,-0.5);
		\draw (-1.1125,0.5) -- (-1.1125,-0.5);
		\draw (-1.1625,0.5) -- (-1.1625,-0.5);
		\draw (-1.2125,0.5) -- (-1.2125,-0.5);
		\draw (-1.2625,0.5) -- (-1.2625,-0.5) node[anchor=north] {$-2^0$};
		\draw (-1.3125,0.5) -- (-1.3125,-0.5);
		\draw (-1.4125,0.5) -- (-1.4125,-0.5);
		\draw (-1.5125,0.5) -- (-1.5125,-0.5);
		\draw (-1.6125,0.5) -- (-1.6125,-0.5);
		\draw (-1.7125,0.5) -- (-1.7125,-0.5);
		\draw (-1.8125,0.5) -- (-1.8125,-0.5);
		\draw (-1.9125,0.5) -- (-1.9125,-0.5);
		\draw (-2.0125,0.5) -- (-2.0125,-0.5);
		\draw (-2.1125,0.5) -- (-2.1125,-0.5);
		\draw (-2.2125,0.5) -- (-2.2125,-0.5);
		\draw (-2.3125,0.5) -- (-2.3125,-0.5);
		\draw (-2.4125,0.5) -- (-2.4125,-0.5);
		\draw (-2.5125,0.5) -- (-2.5125,-0.5);
		\draw (-2.6125,0.5) -- (-2.6125,-0.5);
		\draw (-2.7125,0.5) -- (-2.7125,-0.5);
		\draw (-2.8125,0.5) -- (-2.8125,-0.5) node[anchor=north] {$-2^1$};
		\draw (0.0125,0.5) -- (0.0125,-0.5);
		\draw (0.025,0.5) -- (0.025,-0.5);
		\draw (0.037500,0.5) -- (0.037500,-0.5);
		\draw (0.05,0.5) -- (0.05,-0.5);
		\draw (0.0625,0.5) -- (0.0625,-0.5);
		\draw (0.07500,0.5) -- (0.07500,-0.5);
		\draw (0.08750,0.5) -- (0.08750,-0.5);
		\draw (0.1,0.5) -- (0.1,-0.5);
		\draw (0.1125,0.5) -- (0.1125,-0.5);
		\draw (0.125,0.5) -- (0.125,-0.5);
		\draw (0.1375,0.5) -- (0.1375,-0.5);
		\draw (0.15,0.5) -- (0.15,-0.5);
		\draw (0.1625,0.5) -- (0.1625,-0.5);
		\draw (0.17500,0.5) -- (0.17500,-0.5);
		\draw (0.1875,0.5) -- (0.1875,-0.5);
		\draw (0.2125,0.5) -- (0.2125,-0.5);
		\draw (0.2375,0.5) -- (0.2375,-0.5);
		\draw (0.2625,0.5) -- (0.2625,-0.5);
		\draw (0.2875,0.5) -- (0.2875,-0.5);
		\draw (0.3125,0.5) -- (0.3125,-0.5);
		\draw (0.3375,0.5) -- (0.3375,-0.5);
		\draw (0.36250,0.5) -- (0.36250,-0.5);
		\draw (0.3875,0.5) -- (0.3875,-0.5);
		\draw (0.4125,0.5) -- (0.4125,-0.5);
		\draw (0.4375,0.5) -- (0.4375,-0.5);
		\draw (0.4625,0.5) -- (0.4625,-0.5);
		\draw (0.48750,0.5) -- (0.48750,-0.5);
		\draw (0.5125,0.5) -- (0.5125,-0.5);
		\draw (0.5375,0.5) -- (0.5375,-0.5);
		\draw (0.5625,0.5) -- (0.5625,-0.5) node[anchor=north] {$2^-1$};
		\draw (0.6125,0.5) -- (0.6125,-0.5);
		\draw (0.6625,0.5) -- (0.6625,-0.5);
		\draw (0.7125,0.5) -- (0.7125,-0.5);
		\draw (0.7625,0.5) -- (0.7625,-0.5);
		\draw (0.8125,0.5) -- (0.8125,-0.5);
		\draw (0.8625,0.5) -- (0.8625,-0.5);
		\draw (0.9125,0.5) -- (0.9125,-0.5);
		\draw (0.9625,0.5) -- (0.9625,-0.5);
		\draw (1.0125,0.5) -- (1.0125,-0.5);
		\draw (1.0625,0.5) -- (1.0625,-0.5); 
		\draw (1.1125,0.5) -- (1.1125,-0.5);
		\draw (1.1625,0.5) -- (1.1625,-0.5);
		\draw (1.2125,0.5) -- (1.2125,-0.5);
		\draw (1.2625,0.5) -- (1.2625,-0.5) node[anchor=north] {$2^0$};
		\draw (1.3125,0.5) -- (1.3125,-0.5);
		\draw (1.4125,0.5) -- (1.4125,-0.5);
		\draw (1.5125,0.5) -- (1.5125,-0.5);
		\draw (1.6125,0.5) -- (1.6125,-0.5);
		\draw (1.7125,0.5) -- (1.7125,-0.5);
		\draw (1.8125,0.5) -- (1.8125,-0.5);
		\draw (1.9125,0.5) -- (1.9125,-0.5);
		\draw (2.0125,0.5) -- (2.0125,-0.5);
		\draw (2.1125,0.5) -- (2.1125,-0.5);
		\draw (2.2125,0.5) -- (2.2125,-0.5);
		\draw (2.3125,0.5) -- (2.3125,-0.5);
		\draw (2.4125,0.5) -- (2.4125,-0.5);
		\draw (2.5125,0.5) -- (2.5125,-0.5);
		\draw (2.6125,0.5) -- (2.6125,-0.5);
		\draw (2.7125,0.5) -- (2.7125,-0.5);
		\draw (2.8125,0.5) -- (2.8125,-0.5) node[anchor=north] {$2^1$};
	\end{tikzpicture}
    \caption{Naive representation of $F^{32}$ density within within $\mathbb{R}$.\label{fig:float_density}}
\end{figure}

The figure \ref{fig:float_density} shows a naive representation of $F^{32}$ density within $\mathbb{R}$.

\subsection{Other formats}
\label{section:other-format}
Other formats are defined in the standard.
The table \ref{tab:ieee formats} describes some features of these formats\footnote{Actually, there are also decimal formats described in the IEEE Standard for Floating-Point Arithmetic. Since it is less used, we won't describe it.}.

\begin{table}[ht]
	\begin{center}
		\begin{tabular}{llll}
		    Size &           & $e$ size& $f$ size\\
		    (bits) & Name  & (bits)     & (bits)\\\hline
			16 & Half  & 5 & 10\\ \hline
			32 & Single  & 8 & 23\\ \hline
			64 & Double & 11 & 52\\ \hline
			128 & Quadruple & 15  & 112\\ \hline
			256 & Octuple & 19 & 236\\ \hline
		\end{tabular}
		\caption{Formats of IEEE Standard for Floating-Point Arithmetic}
		\label{tab:ieee formats}
	\end{center}
\end{table}

Number encoding for these formats follows the same rules as for \texttt{float32}.
The only parameter that differs is the size of the bit transposition of the encoded numbers.

\textit{Extended precision} formats are also defined in the standard.
It allows a greater precision than the basic floating point formats.
It uses 40 bits or 80 bits. It does not encode numbers exactly the same way :
it uses also a bit for the integer part.
Since it is rarely used at the code level, we will focus on binary floating-point formats.

\subsection{Mixed precision theory}

Mixed precision corresponds to a calculation method that uses different levels of precision for the same operation.
The aim is to benefit from the shorter calculation time of coarse precision, while maintaining the accuracy of the finer precision. There are several methods of applying the mixed-precision principle.

One of them is to start the calculations start with \texttt{float16} values for rapid matrix math.
But as the numbers are computed, the machine stores the result at a higher precision.
For instance, if multiplying two 16-bit matrices together, the answer is 32 bits in size.
By accumulating the answers, the accuracy becomes finer and finer, until it reaches a level of precision equivalent to those obtained using \texttt{float64}. Since calculations are made with \texttt{float16}, they are faster, less memory is used and power consumption is also lower. 
This operation is called Fused Multiply and Add (FMA). 

The FMA operation compute operations like $x\times y+z$ (where $x$, $y$ and $z$ are floating-point numbers) as a single floating point operation. The classical approach would be to first perform $x\times y$, round, then add the result with $z$, and round again. The FMA calculates $x\times y+z$ at once, then rounds. Thus, the FMA is faster, and more accurate.
On the Itanium processor, the FMA operation requires the same number of cycles as multiplication or addition \cite{graillat2006accurate}.

Since the product of two matrices is equivalent to making sums of products (see equation \ref{matrix_product}), the FMA operation is particularly suitable. The coefficients of the two matrices ($a_{i,k}, b_{k,j}$) can be in \texttt{float16}, while the result ($c_{i,j}$) will be in \texttt{float32}. 

\begin{equation}
	\label{matrix_product}
	\forall (i,j) \in \mathbb{N}_m\times \mathbb{N}_p \quad c_{i,j} = \sum_{k=0}^n a_{i,k}b_{k,j}
\end{equation}
Where $m,n, p \in \mathbb{N}$, $(a_{i,j}) \in M(m,n)$, $(b_{i,j}) \in M(n,p)$, et $(c_{i,j}) \in M(m,p)$ the result of $(c_{i,j}) = (a_{i,j})* (b_{i,j})$.

\section{Main application : machine learning} 

Training a large neural network requires the ability to perform a large number of operations per second.
Recent work on the CIFAR-10 database \cite{krizhevsky2009learning} has reduced the misclassification rate to below two percents \cite{huang2019gpipe, DBLP:journals/corr/abs-1905-01392, DBLP:journals/corr/abs-1805-09501}.
The best classifier uses 557 million parameters.
The top-3 neural networks in this dataset all use at least 10 million parameters.

In the following sub-sections, we will discuss some proposed methods for using different representations during the training phase.
These methods aim to increase the computing speed and decrease the memory used while maintaining the training performance.

\subsection{Using fixed-point numbers with stochastic rounding}

Deep neural networks can be trained using 16-bit fixed-point number instead of \texttt{float32}. 
Some explanation of fixed-point number representation is given in the appendix \ref{appendix:fixed-point numbers}.
By using stochastic rounding, there is hardly any degradation in the classification accuracy \cite{DBLP:journals/corr/GuptaAGN15}. In the following lines we explain the results of some work on this subject.

Using fixed-point numbers requires a conversion stage to go from floating-point representation to its new representation.
Let us denote \texttt{FL} as the number of fractional bits in the fixed point representation, and $\epsilon = 2^{-\texttt{FL}}$ the gap between two numbers.
For a real number $x$, we denote $\lfloor x \rfloor$ as the largest multiple of $\epsilon$ less than or equal to $x$.
We consider the two following rounding scheme:

\begin{itemize}
	\item Round to nearest : the rounding of a given real number $x$ is set to minimize the distance between this number and it's rounding.
	\item Stochastic rounding : the rounding of a given real number $x$ can be either the nearest higher value or the nearest lower value. The probability that its rounding is the value just below is $1-\frac{x-\lfloor x \rfloor}{\epsilon}$.
\end{itemize}
Several trainings were done on the MNIST dataset \cite{lecun1998mnist} using fully connected Deep Neural Networks (DNNs). These trainings compared the two rounding methods by varying the length for fixed-point numbers. The control training is the one using \texttt{float32}. By using stochastic rounding, the loss decreases in the same way for \texttt{float32}, 14-bit, 10-bit and 8-bit fixed-point numbers. It means that the neural network trains just as good, whether it uses \texttt{float32} or 8-bit fixed-point numbers computations. Thus, it allows to use less memory, less power, and compute faster.

\subsection{Using \texttt{float16} instead of \texttt{float32}}

Another way for training deep neural networks is using \texttt{float16} instead of \texttt{float32}. Since \texttt{float16} use half as much memory, the memory requirement can be halved.
However, the use of \texttt{float16} can result in the loss of essential information. To avoid this, and to obtain results as good as those obtained with \texttt{float32}, some methods have been proposed \cite{DBLP:journals/corr/abs-1710-03740}.

\subsubsection{Store \texttt{float32} model to preserve small weights}
While training and updating \texttt{float16} weights during the backward propagation, many weight are bound to become very small. 
The shortest \texttt{float16} possible positive number is $ 2^{-14} \times 2^{-10} = 2^{-24} \approx 5.96 \times 10^{-8}$ (see section \ref{section:other-format}).
If the weight is below, is will be set to $0$.
Small weights have been shown to make a significant contribution to the learning capacity of networks (\cite{DBLP:journals/corr/abs-1710-03740} figure 2a).
One way to prevent the disappearance of the small weights is to store a \texttt{float32} copy of the model.
This copy would accumulate the gradient after each optimizer step.
The results obtained with storage of a copy of the weights prove to be much better than those obtained without storage.
Actually, these results are as good as those obtained with \texttt{float32} only.

\subsubsection{Scalling the loss}

The loss is calculated for each batch to prepare the backpropagation.
Since the main purpose of a classification problem is to minimize the loss, the loss will decrease and reach very low values. During a \texttt{float32} training of Multibox SSD detector, it was observed that the network activation gradient values are mostly below $2^{-24}$\cite{DBLP:journals/corr/LiuAESR15}, which is the smallest value that \texttt{float16} can encode. Since almost no value exceeds $2^{-8}$, scaling up the gradients allows them to occupy more of the \texttt{float16} representable space\footnote{When training neural network, you can have another gradient related issue : gradients can explode when bakpropopagating. This is when they get exponentially large from being multiplied by numbers larger than 1.  One method of preventing this problem is the \textit{gradient clipping}. It will clip the gradients between two numbers to prevent them from getting too large.} Scalling the gradient by a factor a $2^4$ is enough to be obtain the same accuracy as \texttt{float32} training.

\subsubsection{Accumulating partial products into an \texttt{float32} value}

The training of neural networks requires a limited number of different operations. For each operation, it is possible that the result may not be completely accurate. The result will often be the closest number to the result that can be encoded in the chosen standard. A computational inaccuracy is shown in the python command lines below\footnote{Python's standard float type is a \texttt{float64}.}.
 
\begin{pythonenv}
	>>> "
	'0.10000000000000000555'
\end{pythonenv}

The smaller the encoding size, the greater the inaccuracy. To maintain a proper accuracy, some networks need that \texttt{float16} vector dot-product accumulates their partial products into an \texttt{float32} value before storing it into the memory \cite{DBLP:journals/corr/abs-1710-03740}. In doing so, the advantage of using reduced precision is partially lost. However, the latest Graphic Processing Units (GPUs) such as NVIDIA's Tesla natively perform these operations. This allows a considerable increase in speed while maintaining the advantages of working with \texttt{float16} \cite{NVIDIA2017v100}.

\section{Hardware}

Since the calculation units are generally designed to operate on double precision floating point numbers and on integers, the use of mixed precision may not be optimal because of the conversion stage required to go from one precision to another. In the following sections we will detail recent efforts in terms of hardware to adapt the computing units to mixed precision.

\subsection{Matrix computing on GPU}

In its original design, the GPU was sized to quickly manipulate memory to speed up image creation and processing. The resulting buffer memory was intended to be sent to the display device. 

The main difference between them and a Central Process Unit is their highly parallel structure. This makes GPUs much more efficient in terms of operating on data that can be processed in parallel. This is especially the case for matrix calculations. Each component of the matrix resulting from an operation depends only on the coefficients of the matrix argument. They can thus all be calculated in parallel.

\subsection{Half-precision without custom hardware}

\begin{figure}[ht]
    \centering
    \includegraphics[width=0.5\textwidth]{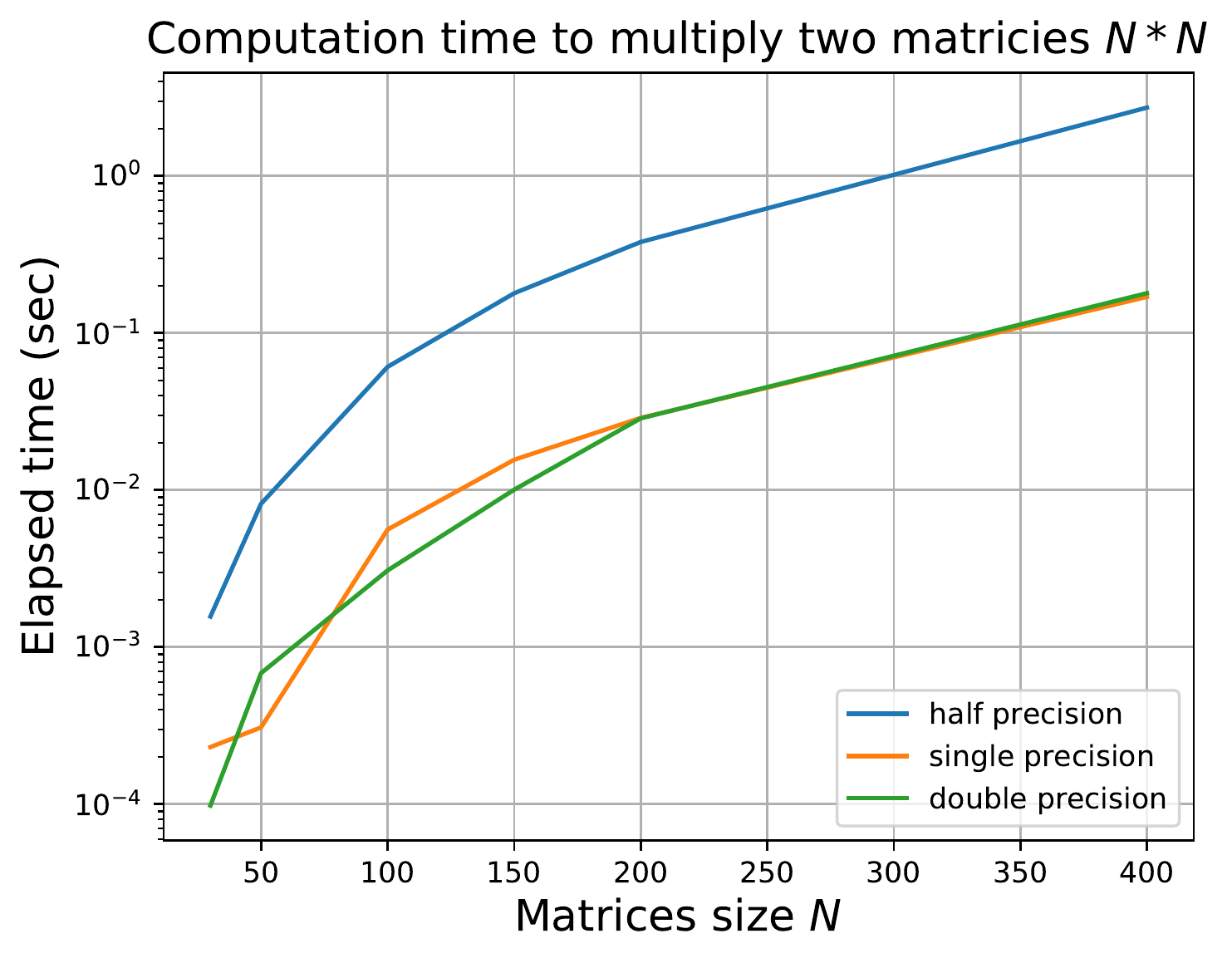}
    \caption{Computation time for a multiplication of two matrices of size $N \times N$ using \texttt{float16}, \texttt{float32} and \texttt{float64}}
    \label{fig:speed compararison}
\end{figure}

We have seen that, with the right architecture, mixed precision can be a very efficient way to increase the computing speed while maintaining the accuracy of \texttt{float32}. 
However, most of the processing units were not designed for \texttt{float16} computation.
In C language, \texttt{float16} does not exist natively.
However, a library can be used to process \texttt{float16} calculations\footnote{Available at : \url{http://half.sourceforge.net/half_8hpp.html}}.
Let us denote $n \in \mathbb{N}$ a natural integer, $A, B\in M_n(\mathbb{R})$.
We want to estimate the time needed to complete the product $A\times B$ for \texttt{float16}, \texttt{float32} and \texttt{float32} format.
The C++ code used, and the main characteristics of the computer used are available in appendix \ref{C_code}.
By varying the size of the matrices, the obtained results are presented in the figure \ref{fig:speed compararison}.

The figure shows a counter intuitive result.
\texttt{float32} and \texttt{float64} need about the same time to compute a product of matrices, irrespective of their size.
What is more, the time needed when using \texttt{float16} is ten times greater than the latter.
The reason for the previous two results is that the processor I used to do these calculations does not natively support \texttt{float16} operations.
The processor converts each input to \texttt{float64}, performs the calculation, and then converts the output back to \texttt{float16} or \texttt{float32} (depending of the entry format) \cite{hyde2003art, techreportintel}.
In some cases, the use of \texttt{float16} can be an advantage. Since the conversion is done directly at the core level, the smaller the size of the data, the greater the capacity to store in cache memory.

\subsection{Tensor core}

The first specialized units using FMA operations to make a product of 4x4 arrays per clock cycle were introduced by the Volta version of NVIDIA GPUs. Using mixed precision, the NVIDIA Tesla V100 accelerator (featuring the Tesla V100 micro-architecture) reaches 125 Tflops/sec. In the following lines, we study how this performance is achieved, and quantify the loss of precision induced by the use of mixed precision.

\begin{figure}[ht]
    \centering
    \includegraphics[width=0.4\textwidth]{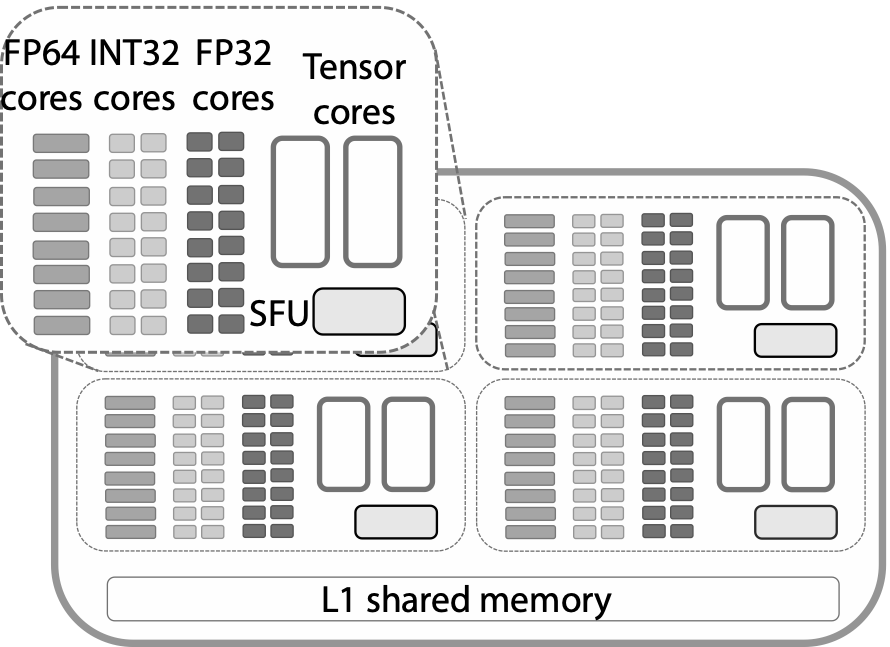}
    \caption{Simplified diagram of the Volta SM architecture. The NVIDIA Tesla V100 uses 80 SMs. Figure from \cite{DBLP:journals/corr/abs-1803-04014}}
    \label{fig:tensor core}
\end{figure}

Figure \ref{fig:tensor core} shows a simplified schematic of the Volta SM architecture. It shows the tensorcore, which comes in addition to the cores dedicated to float and integer operations.

In practice, the tensor cores have been able to deliver up to 83 Tflops/s in mixed precision\footnote{Measurement performed on a Tesla V100 GPU.}. 

\subsection{In-memory mixed precision}

\begin{figure}[ht]
    \centering
    \includegraphics[width=0.5\textwidth]{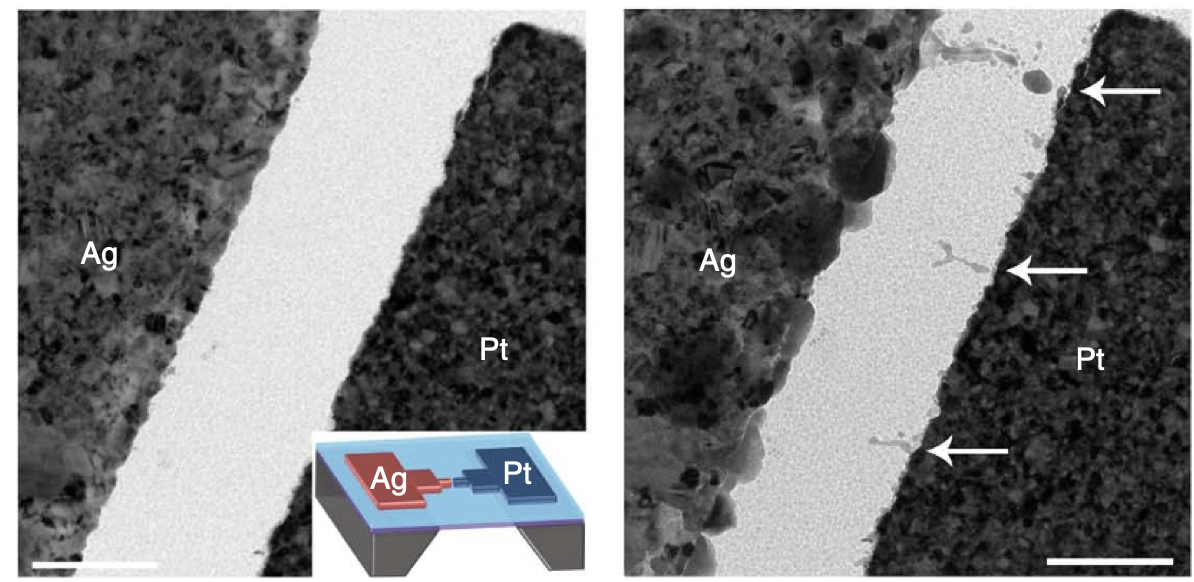}
    \caption{TEm image of an as-fabricated SiO2-based resistive memories. scale bar: 200 nm. Figure from \cite{yang2012observation}}
    \label{fig:resistive-memory}
\end{figure}

Traditional von Neumann architecture reaches its limits. Solutions such as in-memory computing, first introduced in 2012 \cite{180560}, can be used to improve performance in terms of computing power.
The spatial separation between the storage unit and the computing unit is one of the main contributions of computing time.
The cache memory of the processors aims at reducing this distance, by selecting the data that will certainly be needed for a next calculation, and keeping them close to the Arithmetic Logic Unit (ALU).
In-memory goes even further, by processing and storing computational data on the same physical devices organized in a computational memory unit.
It uses nanoscale resistive memory devices within a computational memory unit.
These units are used for both processing and memory. 
The figure \ref{fig:resistive-memory} shows a view of an example of resistive memory.

A hybrid system has been studied, in which a von Neumann machine and a computational memory unit coexist \cite{DBLP:journals/corr/GalloSMMTBCE17}.
The calculation memory unit performs the coarse part of a calculation, and the von Neumann machine implements a backtracking method to iteratively improve the accuracy of the result.
It can be defined as a mixed-precision in-memory computing.
The goal of this hybridization is to combine the high precision of digital computing with the energy efficiency of in-memory computing.

Phase-Change Memory (PCM) are resistive memory devices that can be programmed to get a specific conductance value.
This value is reached by changing the configuration of the amorphous and crystalline phase within the device. It exploit the behaviour of chalcogenide glass.
One way to make the glass amorphous is to change the coordination state of the Germanium atoms with a laser pulse \cite{Simpson2011InterfacialPM}.
One million of these devices have been implemented in a prototype chip. 

\begin{figure}[ht]
    \centering
    \includegraphics[width=0.4\textwidth]{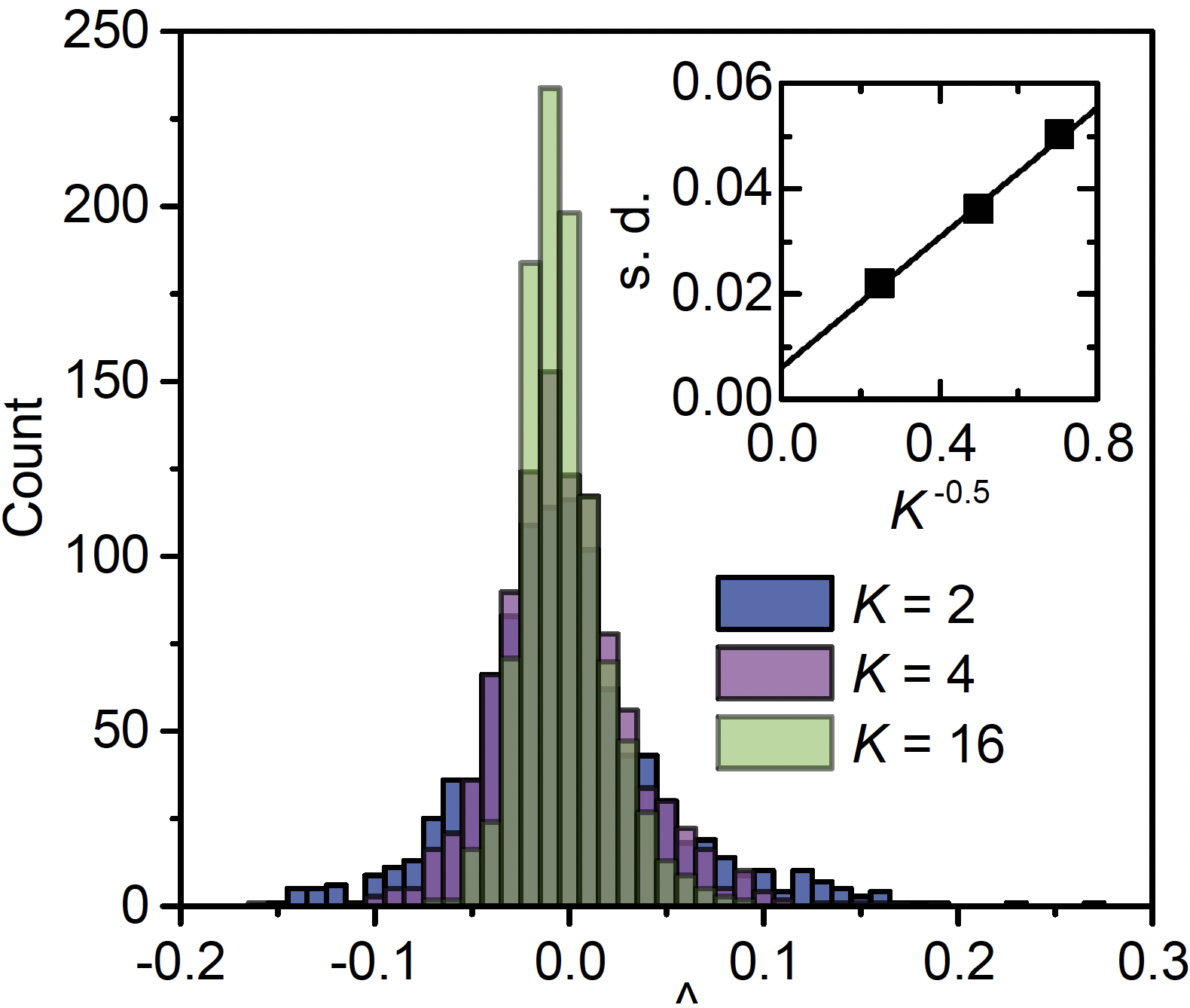}
    \caption{Distribution of scalar multiplication error using $1024 \times K $ PCM devices \cite{Simpson2011InterfacialPM}}
    \label{fig:theta results}
\end{figure}

To test the performance of the latter, the chosen case study is the multiplication of a matrix by a vector. First, let $\beta_n$ , $\gamma_n$ be numbers generated uniformly in $[0,1]$, And $\theta_n = \beta_n\gamma_n$. Let $\hat{\theta}$ be the averaged result on $K$ PCM devices used. The calculation is performed 1024 times. The figure \ref{fig:theta results} shows the error $\hat{\theta}- \theta$ 
distribution with different values of $K$. The standard deviation cis of the order of $K^{-0.5}$, and the mean is $0$.

By accurately solving a system of 5 000 equations using 998,752 CFM devices, these devices have proven their effectiveness in this case of use.

\section{Conclusion and future work}

The loss of accuracy due to mixed precision can be an obstacle to the adoption of this calculation method.
However, we have shown that with some adaptation, the computing algorithm can benefit greatly from mixed precision. In most cases, it is possible to obtain the same level of precision as those obtained with the double precision calculation.
NVIDIA is confident that applications that require a large amount of computing capacity are very likely to benefit greatly from using NVIDIA Tensor Cores and mixed precision. 
A hybrid system comprising a calculation memory unit was imagined and created. It performs the major part of a given calculation task while a processing unit iteratively improves the result.
Today's areas of work for NVIDIA are testing Tensor Cores on applications such as Nek5000 \cite{DBLP:journals/corr/OffermansMSGFSO17} or Fast Multipole Method-accelerated (FFT) \cite{cecka2017low}. Other works on the hardware aim to transpose this in-memory calculation method for applications other than linear system resolution, e.g. Machine Learning.

\section*{Acknowledgement}
I would like to thank my teachers, Ian O'Connor, Laurent Quiquerez and Alberto Bosio who supervised me in the realization of this work.
\bibliographystyle{unsrtnat}  
\bibliography{biblio}
\clearpage
\appendix

\section{Fixed-point numbers}
\label{appendix:fixed-point numbers}

Fixed-point number representation is a data type that represents a finite, fixed number of numbers after the decimal point. 

\[
	\underbrace{\texttt{XXX} \cdots \texttt{XXX}}_{\text{integer part ($m$ bits)}}.
	\underbrace{\texttt{XXX} \cdots \texttt{XXX}}_{\text{fractional part($n$ bits)}}
\]

\noindent
Let's index the binary values like this.
\[
\underset{b_{n+m-1}}{\texttt{X}} \cdots \underset{b_{n}}{\texttt{X}} . \underset{b_{n-1}}{\texttt{X}} \cdots \underset{b_{0}}{\texttt{X}}
\]

Then, the value encoded is :
\begin{equation}
	\displaystyle\sum_{i=0}^{n+m-1} b_{i} 2^{i-n}
\end{equation}

The density of fixed-point numbers is constant in the real numbers. The gap between two values is always $2^{-n}$.

The main advantage of using a fixed-point representation is performance. The value stored in memory is an integer, and the calculation units have very good performance in terms of integer operation.

\section{Comparison of calculation speeds according to the precision used}
\label{C_code}

The code presented in the listing \ref{C code} was used with values of \texttt{N} between 30 and 400 and with \texttt{TYPE} with values \texttt{half}, \texttt{float}, \texttt{double}. To run the code, you will need download the \texttt{half.hpp} header available on \url{http://half.sourceforge.net/half_8hpp.html}. 
The computer used to obtain the results showed in figure \ref{fig:speed compararison} is described in the following lines.\\

\begin{table}[ht]
    \centering
    \begin{tabular}{ll}\hline
        Model &	MacBook Pro\\
    	Model id &	MacBookPro15,4\\
    	Processor Name & Quad-Core Intel Core i5\\
    	Processor Speed & 1,4 GHz\\
    	Number of Processors & 1\\
    	Total Number of Cores & 4\\
    	L2 Cache (per Core) & 256 Ko\\
    	L3 Cache & 6 Mo\\
    	Hyper-Threading Technology & Enabled\\
    	Memory & 8 Go\\
      	Compiler & Apple clang version 11.0.0 (clang-1100.0.33.16)\\
    	Target & x86\_64-apple-darwin19.2.0\\
    	Thread model & posix \\ \hline
    \end{tabular}
    \caption{Technical specifications of the computer used for the experiments}
    \label{tab:my_label}
\end{table}

\begin{lstlisting}[language=C++, caption=Code used to estimate to time needed to perform the multiplication of two matrices \texttt{A} and \texttt{B} of size \texttt{N}, label=C code]
#include "half.hpp"
#include <iostream>
#include <ctime>

using half_float::half;

#define N 150 // Choose matrices size
#define TYPE half // Choose type

int main()
{
  // Define the matrices
  TYPE A[N][N];
  TYPE B[N][N];
  TYPE C[N][N];

  // Define matrices coef bounds
  float LO = -128;
  float HI = 128;

  // Variable used for timing
  clock_t end;
  clock_t begin;
  double elapsed_secs;

  // Set random number in A and B.
  srand((unsigned int)time(NULL));
  for(int i=0; i<N; ++i)
    for(int j=0; j< N; ++j)
    {
      A[i][j]=LO+rand()/(RAND_MAX/(HI-LO));
      B[i][j]=LO+rand()/(RAND_MAX/(HI-LO));
    }

  begin = clock();

  // Multiplying A and B.
  for(int i=0; i<N; ++i)
    for(int j=0; j<N; ++j)
      for(int k=0; k<N; ++k)
      {
        C[i][j] += A[i][k] * B[k][j];
      }

  end = clock();

  // Show the elapsed time
  elapsed_secs = double(end-begin)/CLOCKS_PER_SEC;
  std::cout << "elapsed secs : " << elapsed_secs << std::endl;

  return 0;
}
\end{lstlisting}
\label{listing:id-du-code}

\section{Notations}
\begin{table}[ht]
	\begin{center}
		\begin{tabular}{ll}
		\hline
			$\mathbb{N}$ & Set of natural numbers\\
			$\mathbb{R}$ & Set of real numbers \\
			$\llbracket i, j \rrbracket $ & Set of integers between $i$ and $j$ included\\ 
			$\llbracket i, j \llbracket $ & Set of integers between $i$ included and $j$ excluded\\
			$E^n$ & Cartesian power of a set $E$\\
			$\#E$ & The cardinality of a set $E$\\
			$\mathbb{B}$ & Boole set : $\{0, 1\}$\\ 
			$f(E)$ & Image set of a set $E$ through a function $f$\\
			$\bar{\mathbb{R}}$ & $\mathbb{R} \cup \{ + \infty , - \infty  \} $ \\
			$M_{n,p}(\mathbb{R})$ & Set of real matrices of size $n \times p$ where $n, p \in \mathbb{N}$\\
			$M_n(\mathbb{R})$ & $M_{n,n}(\mathbb{R}) \quad n \in \mathbb{N}$\\
			$\mathcal{F}^{32}$ & Function from $B^{32}$ to  $\bar{\mathbb{R}}$ which gives the real number \\ & associated with its floating point representation. \\
			$F^{32}$ & $\mathcal{F}^{32}(B^{32})$\\
			$F^{32}_{[2^n, 2^{n+1}[}$ & $ F^{32} \cap [2^n, 2^{n+1}]$ : all real numbers that have a \texttt{float32} representation\\\hline
		\end{tabular}
		\caption{Notations}
		\label{tab:notations}
	\end{center}
\end{table}

\end{document}